# Arguing conformance with data protection principles

**Chris Smith**
Department of Computer Science, University of York

**Richard Hawkins**
Department of Computer Science, University of York

*Abstract*—We show how conformance arguments can be used by organisations to substantiate claims of conformance to data protection principles.  Use of conformance arguments can improve the rigour and consistency with which these organisations, supervisory authorities, certification bodies and data subjects can assess the truth of these claims.

Keywords: Data Protection, Privacy Protection, Law, Regulation, Accountability, Assurance, Conformance Arguments

■ **DATA PROTECTION PRINCIPLES** define 'norms for the collection, retention, use and disclosure of personal information' [1].  Such principles underpin legal and self-regulatory policy instruments that are used to regulate the processing of personal data.  By processing personal data, organisations that are subject to these policy instruments are claiming that their processing activities conform to each of these principles.  Claims may be stated explicitly by organisations.  For example, claims may be stated in privacy information published on the organisation's website.  Claims may also be implied simply from the processing of personal data by an organisation, the jurisdiction in which the organisation resides, and therefore the policy instruments to which the organisation is subject.

Claims of conformance with data protection principles may be assessed by organisations that process personal data, supervisory authorities, certification bodies, and data subjects.  Meaningful assessment of conformity requires claims to be substantiated with arguments and evidence that are sufficient for their truth to be determined with confidence.  Arguments and evidence that are not sufficient will lead to assessments of non-conformance (through indetermination) or assessments of conformance that lack rigour and credibility.  However, limited guidance exists to inform organisations on the content, structure and presentation of their arguments and evidence.  Organisations may therefore construct arguments and evidence that are insufficient for meaningful conformity assessment. Additionally, the content, structure and presentation used for arguments and evidence may vary significantly within and between organisations; adding complexity, costs and potential inconsistency to conformity assessment processes.

Conformance with safety standards is essential to safety-critical systems [2] as deviations from desired behaviour can lead to physical harm for individuals. Conformance arguments have been previously proposed as a method by which organisations can substantiate claims of conformance with safety standards [3]. Conformance arguments are structured, comprehensive, and defensible arguments which demonstrate the interpretation of a standard by an organisation. Conformance arguments are supported by evidence that is sufficient to show that a software process or arteiact adequately meets the requirements of the standard.

In this paper, we show how conformance arguments can also be used by organisations to substantiate claims of conformance with data protection principles; improving the rigour and consistency with which conformance can be assessed by organisations, supervisory authorities, certification bodies and data subjects. Conformance arguments could play a crucial role for organisations throughout the system lifecycle. Before deployment, conformance arguments could be used to inform design, implementation and deployment decisions, support engagement with data subjects and underpin both internal and external assessment processes. After deployment, conformance arguments could be used to better understand how system changes may violate conformance with data protection principles. Additionally, should claims of non-conformance be made against an organisation, conformance arguments could be used to defend their decisions.

DATA PROTECTION

We consider data protection to represent privacy as it pertains to the processing of personal data. In this context, violations of privacy arise from processing of personal data that is inconsistent with individual and societal expectations. Violations of privacy can lead to different types of harm [4] for individuals, organisations, groups and society. Individuals may, for example, incur reputational harm from processing of their personal data in a manner that violates their privacy. Organisations may, for example, incur financial harm due to fines imposed from processing of personal data that violates the privacy of individuals to whom it relates. Society may, for example, incur economic harms due to lack of trust and therefore lack of engagement with public systems (eg. health) by individuals whose privacy has been previously violated. Any potential benefits from processing of personal data must therefore be balanced against potential harms from violations of privacy.

Privacy has been acknowledged as an individual [5] and societal [6] value that requires protection. Increased prevalence of systems that process personal data and which have the potential to violate privacy has led to the development of policy instruments at national and international level that aim to define baseline expectations for processing of personal data. Examples of such instruments include those proposed by the Council of Europe [7] and OECD [8]. These instruments have converged around a set of principles that form the basis of legal instruments for data protection in different jurisdictions across the world, including the United Kingdom and European Union. Conformance of processing activities with these principles is monitored and enforced by a supervisory authority with appropriate powers of investigation and enforcement within the particular jurisdiction.

Data protection principles manifest in the UK General Data Protection Regulation (UK GDPR) [9] as follows:

1. Personal data shall be:

    a. processed lawfully, fairly and in a transparent manner in relation to the data subject ('lawfulness, fairness and transparency'); collected for specified, explicit and legitimate purposes and not further processed in a manner that is incompatible with those purposes; further processing for archiving purposes in the public interest, scientific or historical research purposes or statistical purposes shall, in accordance with Article 89(1), not be considered to be incompatible with the initial purposes ('purpose limitation');

    b. adequate, relevant and limited to what is necessary in relation to the purposes for which they are processed ('data minimisation');

    c. accurate and, where necessary, kept up to date; every reasonable step must be taken to ensure that personal data that are inaccurate, having regard to the purposes for which they are processed, are erased or rectified without delay ('accuracy');

    d. kept in a form which permits identification of data subjects for no longer than is necessary for the purposes for which the personal data are processed; personal data may be stored for longer periods insofar as the personal data will be processed solely for archiving purposes in the public interest, scientific or historical research purposes or statistical purposes in accordance with Article 89(1) subject to implementation of the appropriate technical and organisational measures required by this Regulation in order to safeguard the rights and freedoms of the data subject ('storage limitation');

    e. processed in a manner that ensures appropriate security of the personal data, including protection against unauthorised or unlawful processing and against accidental loss, destruction or damage, using appropriate technical or organisational measures ('integrity and confidentiality').

2. The controller shall be responsible for, and be able to demonstrate compliance with, paragraph 1 ('accountability').

Conformance of processing activities with these principles by organisations in the UK, or by organisations that are processing the personal data of individuals in the UK, is monitored and enforced by the Information Commissioner's Office (ICO). The ICO is the supervisory authority for data protection legislation in the UK and has investigative and enforcement powers that are also defined in the legislation.

Conformance of processing activities with data protection principles is the basis of data protection by design and default (DPbD). DPbD is itself an obligation for organisations under Article 25 of the UK GDPR. DPbD requires organisations to 'implement appropriate technical and organisational measures' [9] to implement the data protection principles effectively, safeguard the rights of data subjects, and ensure only the data necessary for each purpose are processed. Factors that determine whether measures are appropriate include: current state of the art; cost of implementation; nature, scope, context and purposes of processing; and risks for rights and freedoms of individuals. Importantly, these factors must be continually assessed as systems progress through their lifecycle.

Demonstration of conformance with data protection principles and, by extension DPbD, is underpinned by the accountability principle. The accountability principle requires that processing undertaken by organisations conforms to the data protection principles, and that organisations are able to demonstrate compliance. We choose to refer to conformance rather than compliance in this work as we consider assessments undertaken not only by supervisory authorities, but also by organisations themselves, certification bodies and data subjects.

Guidance published by the ICO [10, 11] outlines their expectations regarding measures through which conformance with data protection principles can be demonstrated by organisations within a data protection audit. These expectations may also be used by organisations to inform internal conformity assessments undertaken for the purposes of internal quality control and improvement, or to prepare for external conformity assessment. We include below an extract from the Policies and Procedures section of guidance published by the ICO in their Accountability Framework [10]:

> ...
> Your organisation's approach to implementing the data protection principles and safeguarding individuals' rights, such as data minimisation, pseudonymisation and purpose limitation, is set out in policies and procedures
> ...

Published guidance provides high-level expectations regarding the evidence needed to substantiate any claims of conformance with data protection principles. Organisations must determine how to construct such evidence and the arguments that coherently tie this evidence to the implementation of the data protection principles. Organisations may choose to pursue certification through an approved certification scheme, or membership of approved sector-specific codes of conduct, to subject their processing activities to external assessments of conformity based on more granular expectations that have been approved by the supervisory authority. However, in the UK, few certification schemes are currently listed on the register of approved certification schemes published by the ICO [12], and those schemes that have been approved have limitations to their applicability.

CONFORMANCE ARGUMENTS

Conformance with safety standards is essential to safety-critical systems [2] as deviations from desired behaviour can lead to physical harm for individuals. Safety standards guide or constrain the development processes, prescribe or proscribe product features and dictate assessment practices. Safety standards help to establish a consistent benchmark against which projects can be measured, minimum standards and best practices can be defined, and maturity in system development and assessment can be improved. However, demonstrating conformance to safety standards is not always straightforward. Safety standards often require interpretation because: i) the standard sets high-level goals rather than specific and prescriptive requirements, ii) the standard contains deliberate non-specificity (to support wider applicability) and iii) it is possible to meet the letter but not the spirit of the standard. Quality, transparency and scrutability of documentation can also vary significantly, which results in a lack of clarity as to exactly what a conformance claim signifies.

Conformance arguments were proposed by Graydon et. al. [3] to provide an explicit, rigorous and structured approach to the generation and communication of evidence relating to claims of conformance with safety standards. Conformance arguments are structured, comprehensive, and defensible arguments which demonstrate that supplied evidence is sufficient to show that an artefact adequately meets the requirements of a safety standard. Assessments of conformity evaluate the validity of the argument structure and the validity of the evidence used to substantiate claims to determine the validity of the overall claim. Conformance arguments can also include dialectic challenges to represent scenarios in which the validity of an existing conformance argument is challenged by system or wider contextual changes. Proposed benefits of conformance arguments for organisations developing safety-critical systems, and for authorities tasked with asserting, assuring and regulating conformance, include: explicitness, transparency, confidence, design efficiency, predictability and repeatability.

Safety-critical systems commonly create a safety argument as part of a safety case, which demonstrates that a system is sufficiently safe to operate. Safety arguments decompose a safety claim into a series of subclaims until these can be solved with evidence. We can use the same approaches used for safety arguments to construct conformance arguments. Conformance arguments, like safety arguments, are rigorous, but necessarily informal, logical arguments. Safety arguments can be recorded in any suitably expressive argument notation. However, it has been generally found to be easier to use a graphical argument notation, such as the Goal Structuring Notation (GSN) [13]. GSN can also be used to represent conformance arguments.

ARGUING CONFORMANCE WITH DATA PROTECTION PRINCIPLES

Conformance arguments have the potential to be used to substantiate claims of conformance with data protection principles. Use of conformance arguments could improve the rigour and consistency with which organisations that process personal data, supervisory authorities, certification bodies and data subjects can assess conformance with data protection principles.

To demonstrate this potential, we outline a hypothetical case study system in which personal data is processed. We show how a conformance argument can be developed to substantiate a claim of conformance with a specific data protection principle for a specific processing activity within this system. We also show how changes in: i) the data that is processed, and ii) the purpose for which the data is processed, can be represented as 'challenges' that may invalidate the conformance argument and require resolution through amendments to arguments, evidence and/or the system itself.

Case Study

**Background**

We consider a hypothetical attendance monitoring system (StudentCheck), which is proposed for development and use in an academic institution. The institution delivers teaching sessions to students who are registered on programmes of study. Students are required to attend teaching sessions associated with modules on their registered programme of study. The institution is considering the development and use of StudentCheck to identify those students who did not attend a required session and to contact these students to provide appropriate support.

**System design**

Figure 1 summarises a proposed design for StudentCheck using a data flow diagram. The diagram includes high-level data flows between StudentCheck and external entities, which include both systems and users. Selected data flows between external entities have also been included to provide additional context.

In preparation for delivery of a module, the Timetabling System sends the details of a module to StudentCheck. Details supplied include the schedule for the module sessions and a list of the students that are registered to the module. Prior to each teaching session, the Facilitator requests a session code from StudentCheck and distributes this code to Students who are present in the session. Students then submit this session code back to StudentCheck to confirm their attendance. The Support Team uses StudentCheck to request details of students who were absent from a given session and send a support email to those students.

Internally, StudentCheck comprises a number of interacting sub-processes. For brevity, we focus only on a specific sub-process: Update Attendance. This sub-process is responsible for processing the session code that is submitted by a Student and updating their session attendance accordingly. We consider this sub-process to be a single processing activity and therefore a single unit of analysis for the purposes of arguing conformance with the data protection principles. Decisions regarding the appropriate granularity at which to model and analyse processing activities within a system are central to the construction of conformance arguments, and to the demonstration of conformance more generally. Figure 2 summarises a proposed design for the Update Attendance processing activity using a data flow diagram.

Following attendance at a session, a Student supplies their user credentials, which comprises their username and password, and the session code to StudentCheck. The Authentication process uses the User Database to confirm the identity of the Student based on the supplied credentials. The Session Management process then uses the Session Database to determine the details of the session based on the supplied session code. We consider the session details to include: a unique session identifier; module identifier; session start timestamp; and session end timestamp. The Attendance Management process uses the user's identity and the session details to update the Attendance Database and then supplies a confirmation of attendance to the Student. To address the potential for session codes to be distributed to, and submitted by Students who have not attended a session, the Attendance Management process imposes a constraint on the time period within which a student's attendance at a session can be confirmed.

**Claims of conformance**

StudentCheck is proposed for development and use by an academic institution in the United Kingdom. Any processing of personal data would therefore engage the data protection regime in the United Kingdom. Any claims of conformance of processing activities would be with respect to the data protection principles as they manifest in the UK GDPR.

For brevity, we focus only on claims of conformance with the data protection principles relating to the Update Attendance activity (see Figure 2). We consider the data that is processed by the Update Attendance activity to fall into the definition of 'personal data', as articulated in the Article 4(1) of the UK GDPR, and to therefore engage the data protection regime. Specifically, we consider that a Student can be identified by reference to the username that is supplied with their user credentials. Additionally, we further restrict our focus to a claim of conformance relating to the data minimisation principle:

*Processing of personal data for the Update Attendance activity conforms to the data minimisation principle of the UK General Data Protection Regulation*

Conformance Arguments

We use three scenarios to demonstrate the potential of conformance arguments in substantiating claims of conformance with the data protection principles. Firstly, we outline a conformance argument for a claim of conformance with the data minimisation principle by the Update Attendance activity. We then introduce challenges to the validity of this conformance argument to reflect: i) changes in the data that is processed by the activity, and ii) changes in the purpose for which data is processed by the activity. We illustrate each scenario using conformance argument fragments expressed in GSN.

**Scenario 1: Constructing a conformance argument for a claim of conformance with the data minimisation principle**

Figure 3 illustrates a conformance argument fragment that was constructed to substantiate the claim relating to conformance of the Update Attendance activity with the data minimisation principle.

The overall claim of conformance is the top-level 'goal' of the argument (G1). Context elements are used to support interpretation and explicitly define the scope for which the claim is valid. C1 references a description of the nature, scope, context and purposes of the processing activity. These dimensions are used within the UK GDPR to characterise processing activities and to inform risk-based decision making in relation to these activities. C2 references a technical description of the personal data that is processed, which might include, for example, a data model or schema. C3 references the data minimisation principle within the UK GDPR. C4 references the concepts of data protection by design and default within the UK GDPR. Provisions within the legislation relating to DPbD provide further information on the different factors that determine whether technical and organisational measures in place to implement data protection principles are appropriate.

We define an argument strategy (S1) that states how the claim of conformance (G1) will be argued. Based on the wording of the data minimisation principle in the UK GDPR, we can decompose our argument into three parts. Conformance can be individually argued for these three parts to collectively substantiate our overall claim (G1). We create three 'sub-goals' (G1, G2, and G3) to represent these parts and construct arguments to achieve each of these sub-goals. Construction of arguments for the sub-goals requires us to interpret the referenced terms: 'adequate', 'relevant' and 'limited to what is necessary in relation to the purposes', and to demonstrate that processing is consistent with this interpretation. We include below an extract from guidance published by the ICO [14], which we used to inform our interpretation of these terms:

> You must ensure the personal data you are processing is:
> 
> 1. adequate – sufficient to properly fulfil your stated purpose;
> 2. relevant – has a rational link to that purpose;
> 3. limited to what is necessary – you do not hold more than you need for that purpose;

It is important to note that interpretation of this guidance is itself subjective. For example, what is meant by "sufficient", "properly" and "rational" requires interpretation. Use of conformance arguments helps to demonstrate how particular concepts have been interpreted, why a particular approach to conformance has been taken, and how sufficiency is justified.

To demonstrate that the personal data processed by Update Attendance is 'sufficient to properly fulfil the stated purpose', the argument must show how the data that is processed is sufficient for the purpose of 'Attendance Monitoring'. For example, if the data processed were to omit any data item(s) by which a student could be uniquely referenced, then it would be difficult to argue that the data is adequate for the purpose. To demonstrate that the personal data 'has a rational link to that purpose', the argument must show that the data which is processed is semantically related to the purpose. For example, if the data processed were to include any data item(s) relating to clubs and societies for which the student is a member, then it would be difficult to argue that the data is semantically related to the purpose. Finally, to demonstrate that 'you do not hold more than you need for that purpose', the argument must show that any data that is processed is necessary for the purpose. For example, if the data processed were to include data item(s) relating to sessions in the previous academic year, then it would be difficult to argue that more data than needed for the purpose is not being held.

**Figure 4** illustrates a conformance argument fragment for the sub-goal relating to adequacy (G2). The goal is associated with context (C5) that supports interpretation of the argument by referencing the description of adequate provided in ICO guidance. Based on the ICO guidance, we define an argument strategy (S2) which states that conformance will be argued by reference to the sufficiency of the data. We associate this strategy with context (C6) which states that sufficiency will be expressed with respect to three dimensions: 1) data subjects to which the data items relate; 2) data items processed for these data subjects; and 3) values of these data items. We decompose the argument into sub-goals relating to these three dimensions (G5, G6, and G7). We also illustrate the decomposition of G5 into two further claims, or sub-goals, which are intended to substantiate the claim in G5.

Different forms of evidence may be used to substantiate claims, including policies and procedures, automated test results and formal proofs. In Figure 4, we reference a justification report (Sn1) and an audit report (Sn2) that might be used to substantiate G8 and G9 respectively.

The justification report would outline the categories of data subjects for whom data items are processed (eg. Registered Students), and provide a justification for why the processing of data items relating to these subjects is sufficient for Attendance Monitoring. The audit report would provide an assessment (ideally independent) that the justification report is sensible. We have not produced the justification or audit reports that would be required by the argument above, or other related arguments, as part of this work. We also have not explored the different forms of reports, or indeed the different forms of evidence that might be used to substantiate G5 or other claims. For brevity, we do not include any evidence nodes for G6 and G7 in Figure 4. However, we would expect that any forms of evidence that are applicable to G5 (and G8 and G9) would also be applicable to G6 and G7.

Data protection regimes and their interpretation by different stakeholder groups, including supervisory authorities, may be subject to changes over time. Conformance arguments should therefore be subject to periodic review throughout the system lifecycle to ensure continued validity. Additionally, processing activities that constitute the system may be subject to changes over time. For example, due to new features and bug fixes. Arguments that were previously constructed may therefore cease to be valid and sufficient to substantiate claims of conformance. Conformance arguments can incorporate dialectic challenges to explicitly represent and address changes to the system and to the policy instruments to which the organisation responsible for the system is subject.

We now briefly explore two scenarios in which dialectic challenges could be used to provide explicit reasoning about the ongoing validity of a conformance argument following a change.

**Scenario 2: Challenging a conformance argument following changes to the data that is processed**

Figure 5 illustrates a conformance argument fragment that includes a challenge to the validity of the previous conformance argument due to proposed changes in the data that is processed by the Update Attendance activity. Such changes may be driven by a range of functional and non-functional system requirements. For simplicity, we consider that user credentials, which includes username and password, have been proposed for omission from the data items processed by the activity in order to fulfil security requirements. Data items that are processed may therefore no longer be sufficient to fulfil the purpose of Attendance Monitoring. A dialectic challenge (CG1) to the claim that the data items are sufficient (G6) can be used to explicitly represent and explore this scenario and the implications for conformance with the data minimisation principle.

Challenges that are included in an argument require an appropriate response to ensure continued validity of the argument. Responses could include rebuttal of the challenge (explaining why the challenge put forward is invalid); making changes to the system design, implementation and/or deployment; and inclusion of additional arguments [15]. We do not explore the use of these potential strategies, or how they might apply to the conformance arguments above. We seek only to highlight how changes in the data that is processed may invalidate a conformance argument, how this potential invalidity can be represented using a dialectic challenge, how different strategies are available to address such challenges, and how pursuit of these strategies should prompt resolution through amendments to arguments, evidence and/or to the system itself.

One of the advantages of conformance arguments is how they can enable more effective identification of possible challenges and prompt discussion and selection of appropriate responses.

**Scenario 3: Challenging a conformance argument following changes in the purpose for which data is processed**

Figure 6 illustrates a conformance argument fragment that includes a challenge to the validity of the previous conformance argument due to changes in the purposes for which data is processed by the activity. Such changes may again be driven by a range of functional and non-functional system requirements, but we consider it more likely that such changes would be driven by functional system requirements. For simplicity, we consider that the purposes for which data is processed has been extended to include Attendance Prediction. A dialectic challenge (CG2) to the context that states the purposes of the processing activity (C1) can be used to explicitly represent and explore this scenario and the implications for conformance with the data minimisation principle. The challenge asserts that the change to purposes means that the sub-claims and associated evidence whose interpretation is contingent on purpose, and the conformance argument as a whole, may no longer be valid.

As outlined in Scenario 2, we have a number of options for responding to changes. For example, one might attempt to construct an argument to assert that processing for 'Attendance Prediction' is 'compatible' with the original purpose of 'Attendance Monitoring'. However, we again do not explore any such strategies or how they might apply to the conformance arguments above.

No arguments or evidence relating to the validity of purposes are included in the conformance argument. For example, claims relating to the legitimacy and specificity of the purposes. However, such arguments and evidence would be required when constructing arguments for real-world scenarios. Such arguments and evidence may, for example, be included in a conformance argument for the Update Attendance activity that pertains to the purpose limitation principle. Conformance arguments relating to different principles for a given processing activity are therefore likely to be interdependent. Such interdependencies mean that challenges to a conformance argument for one principle (eg. purpose limitation) have the potential to also invalidate argument for other principles (eg. data minimisation).

DISCUSSION

Limitations

We outlined a conformance argument for a single data protection principle (Data Minimisation) for a specific processing activity (Update Attendance) within a hypothetical case study system (StudentCheck) at a specific stage of the system lifecycle (Design). Construction of arguments relating to other data protection principles for processing activities in different systems at different stages in the system lifecycle would highlight additional considerations. We also focused solely on data protection principles as they manifest within the data protection regime in the UK, drawing on legislation and guidance specific to the UK.

Arguments and supporting evidence presented were intended to be indicative only. Where challenges to conformance arguments were presented, we did not explore the different options that might be used to address these challenges. We also do not compare and contrast the use of conformance arguments with existing practices that are used by organisations to demonstrate conformance with data protection principles.

Benefits

Construction of conformance arguments enables organisations to develop increased confidence regarding the truth of their claims of conformance with data protection principles at all stages of the system lifecycle. Organisations can use arguments to inform design, implementation and deployment decisions, support engagement with data subjects and underpin both internal and external assessment processes. Conformance arguments also have the potential to benefit those stakeholders responsible for external conformity assessments, including supervisory authorities and certification bodies, by increasing the consistency and rigour of arguments and evidence across organisations, and reducing costs.

Construction and evaluation of conformance arguments is also likely to prompt the development of knowledge and skills relating to the theory and practice of data protection across stakeholders; supporting more informed decisions about processing of personal data more generally. Construction and evaluation of conformance arguments may also highlight challenges that are faced by organisations in operationalising and arguing conformance with data protection principles; potentially driving updates to data protection legislation, and guidance published by supervisory authorities and certification bodies.

Challenges

Construction and evaluation of conformance arguments is challenging, particularly for individuals and teams without prior experience. Organisations require knowledge of all policy instruments that are engaged by a system, knowledge of relevant technical details of the system, and the skills to effectively integrate and synthesise this knowledge into robust conformance arguments. Organisations are likely to require input from a multi-disciplinary team whose expertise spans areas including Computer Science, Law and Ethics. Appropriate recruitment and up-skilling is therefore likely to be required across stakeholders to support a conformance argument approach.

Organisations must also be able to determine the appropriate level of granularity at which to model the system and construct the conformance argument(s); balancing the rigor of the resultant arguments with the resources required to construct and maintain the argument.

Adoption of conformance arguments would also be subject to interdependencies between stakeholder groups, such that adoption by one group, for example, organisations, would be dependent on adoption by another group, for example, supervisory authorities. Early stages of adoption are likely to require significant engagement and collaboration between stakeholder groups to ensure convergence on mutually consistent practices.

Future work

Due to the limited guidance that exists to inform organisations on content, structure and presentation, we have suggested that organisations may construct arguments and evidence that are inconsistent, and insufficient for meaningful conformity assessment. However, additional work is required to identify and characterise the practices that are currently used by organisations to demonstrate conformance, and to determine the extent of divergence in these practices within and between organisations.

Additional work is required to determine whether the use of conformance arguments would be sufficient to underpin meaningful assessment of conformity by, for example, supervisory authorities, and to explore barriers to adoption for different stakeholder groups. Qualitative studies exploring more detailed (and potentially real-world) case studies with different stakeholder groups would help to highlight specific opportunities and challenges.

Adoption of conformance arguments is likely to require education and training to be developed and delivered across stakeholder groups. Organisations would benefit from the step-by-step guides, standard operating procedures, and templates that define widely applicable argument and evidence structures.

Future work might also explore those forms of arguments and evidence that are most appropriate to substantiate claims relating to specific data protection principles at different stages in the system lifecycle.

Generic tools exist to support construction of arguments expressed in GSN. However, tools that are specific to data protection could provide further assistance for those constructing and evaluating arguments. Tools might also be developed to highlight post-deployment changes to processing activities, such as component updates, and enable re-evaluation of arguments. Conformance arguments are likely to be managed as 'living' artifacts, which are updated throughout the system lifecycle. Tools and methods that support ongoing management of these artefacts are likely to be helpful for organisations and other stakeholder groups responsible for conformity assessment.

CONCLUSION

We have shown how conformance arguments can be used to substantiate claims of conformance with data protection principles;

improving the rigour and consistency with which organisations, supervisory authorities, certification bodies and data subjects can assess conformance with data protection principles. Our work has also highlighted some challenges associated with the adoption of this approach, including the expertise and time required to construct, maintain and evaluate arguments. Future work to support this approach should focus on engagement, education and training programmes for the different stakeholder groups in the data protection ecosystem, and on tools to improve the effectiveness and efficiency with which arguments and evidence can be constructed, maintained and evaluated.

ACKNOWLEDGEMENTS

**Chris Smith** was a Lecturer in the Department of Computer Science at University of York in the United Kingdom. His research interests include privacy protection, accountability and trustworthiness. Smith received his PhD in Computing Science from Newcastle University. Contact him on LinkedIn at: https://www.linkedin.com/in/drcjsmith/.

**Richard Hawkins** is an Associate Professor at the University of York in the United Kingdom. His research interests include safety cases and safety assurance for autonomous systems and AI. Hawkins received his PhD in Computer Science from the University of York. Contact him at: richard.hawkins@york.ac.uk.

APPENDIX A

Design of the StudentCheck system was inspired by the use of a system at the University of York[1]. In particular, the design uses 'session codes' as the means by which attendance at sessions is indicated. We speculated on the specific processing activities that might underpin such a system and the design that is presented is the result of this speculation. We do not assert any correspondence between the speculated activities and those that comprise the system in use at the University of York, or any other similar system. We also do not assert that the speculated activities represent all of those that would be required for such a system. We do however believe that the speculated activities are sufficiently realistic to demonstrate the applicability of conformance arguments to claims relating to data protection principles

---

[1] For further information, see:
https://www.york.ac.uk/students/support/check-in

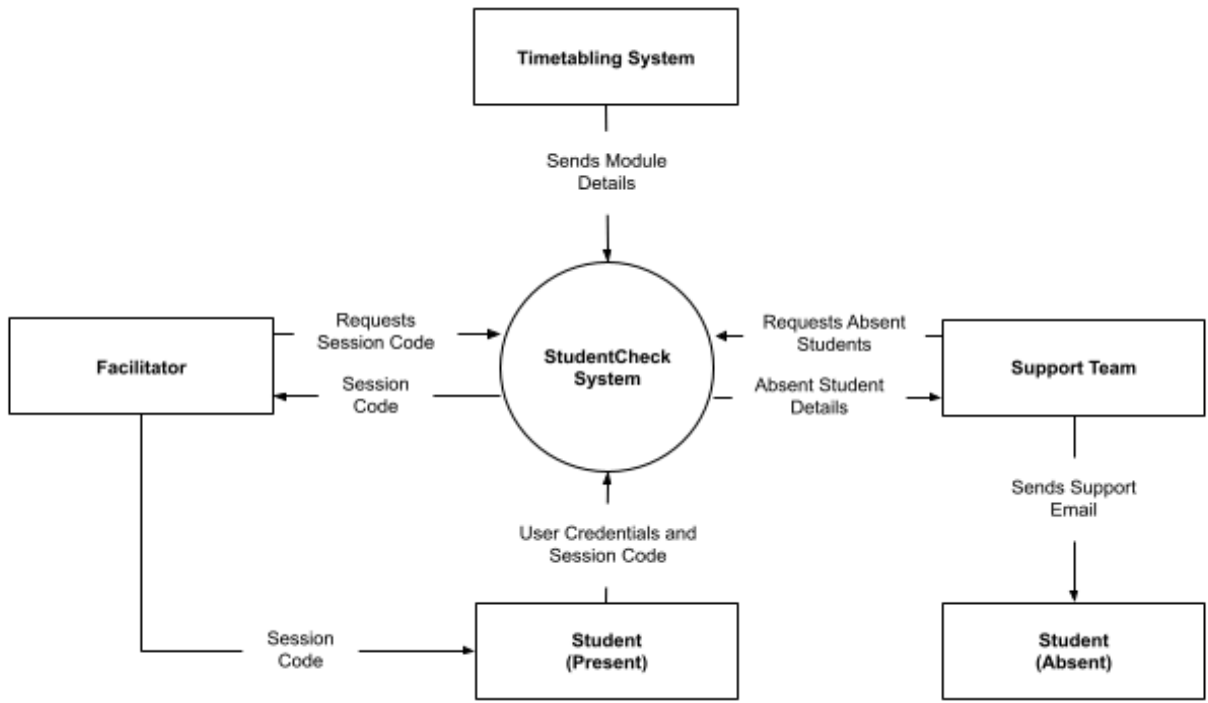

Figure 1. Data Flow Diagram for the StudentCheck System

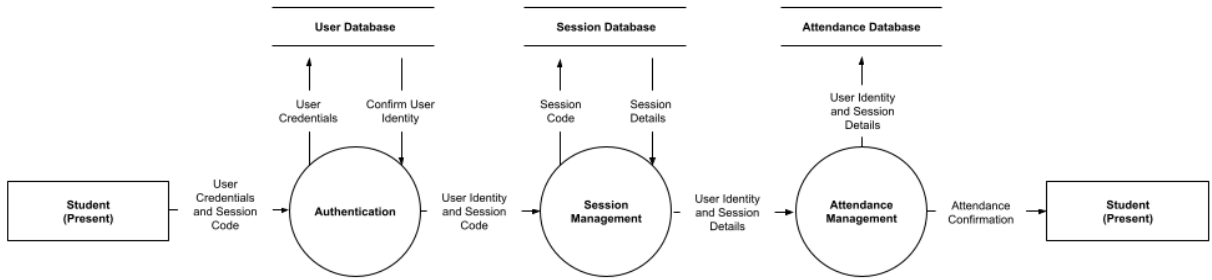

Figure 2. Data Flow Diagram for the Update Attendance processing activity

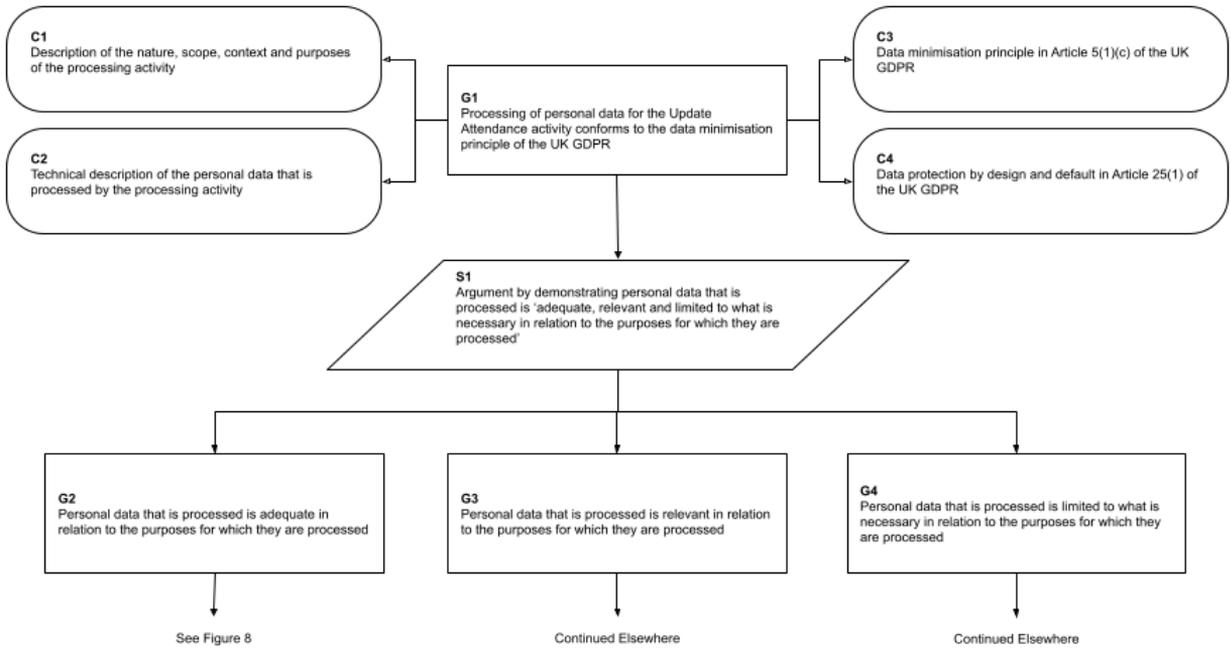

Figure 3. Conformance argument fragment in GSN for the claim that the Update Attendance processing activity conforms to the data minimisation principle of the UK GDPR

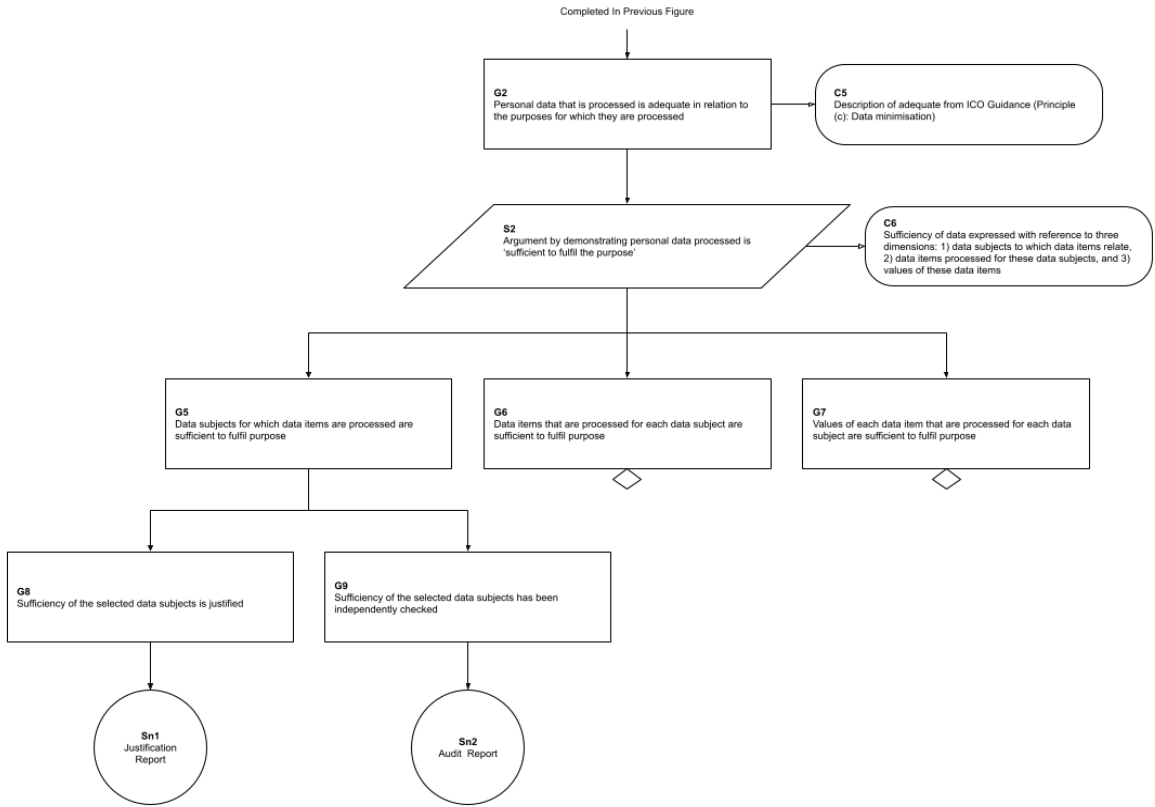

Figure 4. Conformance argument fragment in GSN for the claim that the personal data that is processed for the Update Attendance processing activity is adequate for the purposes for which they are processed

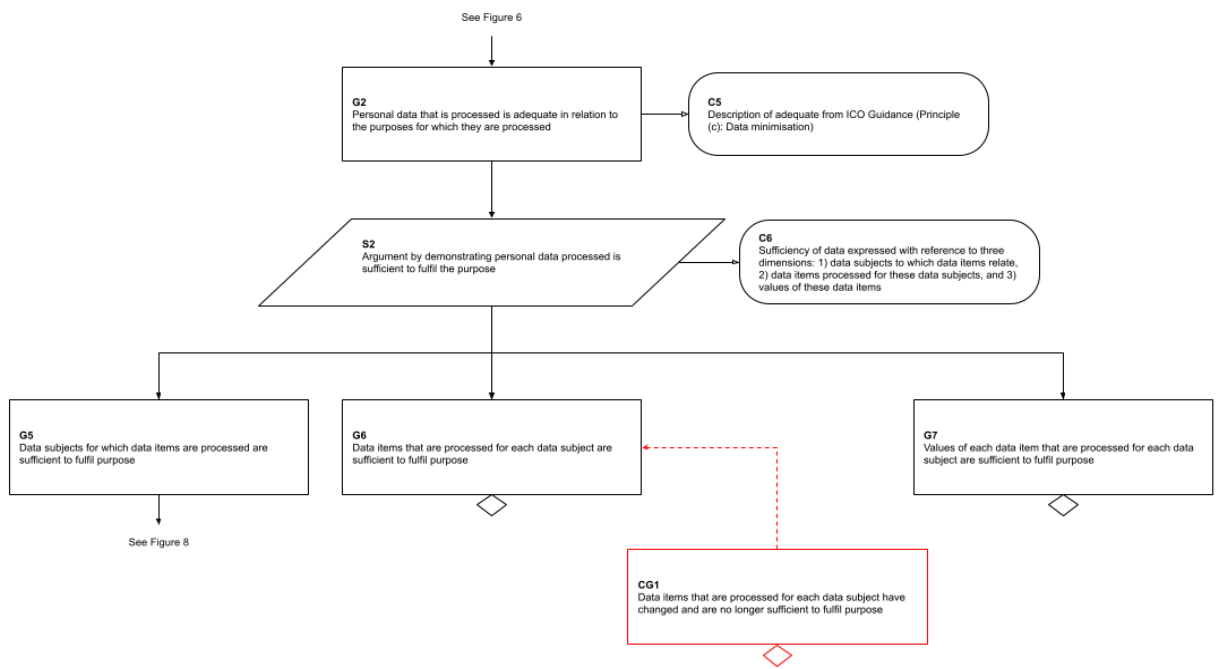

Figure 5. Conformance argument fragment in GSN which includes a dialectic challenge to the claim that personal data processed for Update Attendance is adequate for the purposes for which they are processed

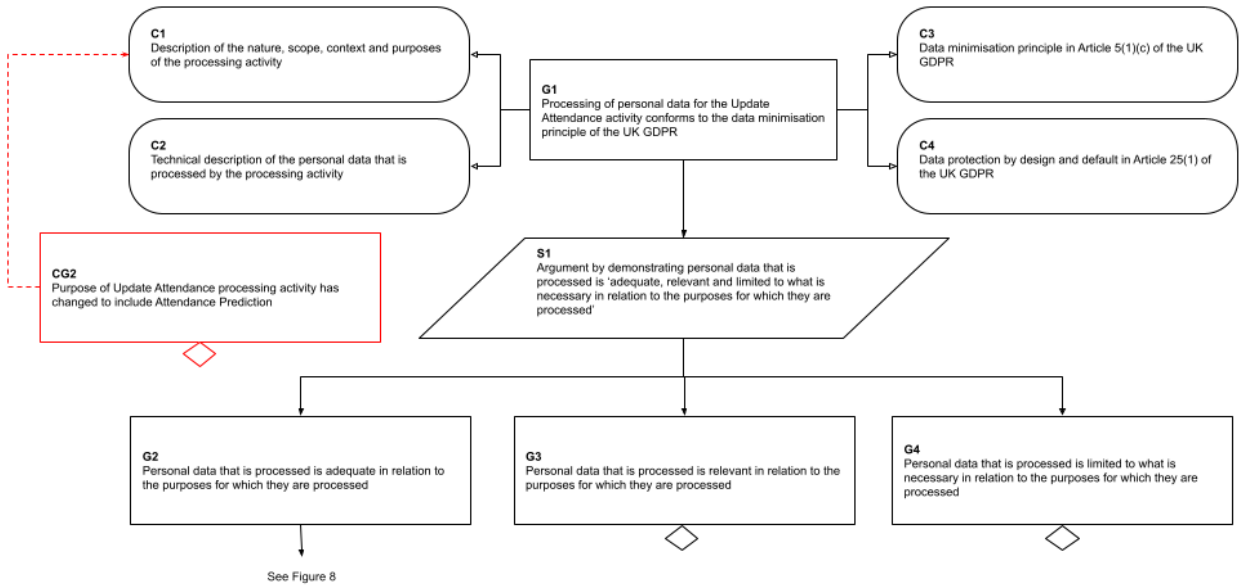

Figure 6. Conformance argument fragment in GSN which includes a dialectic challenge to the purposes for which personal data is processed by the Update Attendance activity